\documentstyle[12pt]{article}
\normalbaselines
\begin{document}

\begin{center}
{\large \bf
ON THE NATURE OF CHARGE CONJUGATION IN QUANTUM THEORY}\\[2mm]
G.A. Kotel'nikov\\{\it Russian Research Center "Kurchatov Institute", 
Moscow 123182, Russia}\\
e-mail: kga@kga.kiae.su\\[3mm]
\end{center}

\begin{abstract}
    On the  basis  of  the  invariance  of  Dirac  equation $(\gamma p
-mc)\Psi (x,c)=0$ with respect to the inversion of the speed of  light
$Q$:   $x\to  x$,  $c\to  -c$,  it  is  shown  that  the  relationship
$[C,PTQ]\Psi(x,c)=0$  between  the  transformations  of   the   charge
conjugation  $'$,  the space inversion $P$,  the time reversal $T$ and
the  inversion  of  the  speed  of  light  $Q$  is  true.  The  charge
conjugation in quantum theory may be interpreted as the consequence of
the discrete  symmetries  reflecting  the  fundamental  properties  of
space, time and speed of light.
\end{abstract}

\section{Introduction}
\label{s1}
The charge conjugation $C$, the space inversion $P$, the time reversal
$T$ and the $CPT$-theorem connected with them play the important  role
in quantum theory \cite{Sch57},  \cite{Ber68}. Their significance, for
example,  in the theory of beta-decay,  was numerously  discussed.  As
this  takes  place,  the $P ({\bf x}\to-{\bf x})$ and the $T (t\to-t)$
symmetries  are  interpreted  as  the  evidence  of  the   fundamental
properties of space and time, namely the equivalence of right and left
directions in space and the possibility of sign reversal of the  time.
The  nature of the charge conjugation is not established \cite{Shi72}.
In the present work we try to connect the charge conjugation $C$  with
the  inversion of the speed of light $Q:  c\to -c$ \cite{Kot92}.  \par
The source of the $Q$-symmetry seems to be the invariance of the light
cone equation relative to the replacement $c\to -c$,  $t\to t$,  ${\bf
x}\to {\bf x}$:
\begin{equation}
\label{f1}
Q: \ c^2{t^2} -{\bf x}^2 = 0 \to (-c)^2{t^2} -{\bf x}^2 = 0
\end {equation}
\noindent It  is  shown in the work \cite{Kot92} that the $Q$-symmetry
is also inherent in the D'Alembert equation,  the  Maxwell  equations,
the  equation  of  movement  of \ a charge particle in electromagnetic
field, that is, the equations of the classical electrodynamics.
\par For  finding  the interrelation between the inversion $Q$ and the
charge conjugation $C$ in quantum theory we use the Dirac equation.

\section{$P$, $T$, $Q$ -symmetry of Dirac equation}
\label{s2}
Let us introduce the Dirac equation in the form [1,2]
\begin{equation}
\label{f2}
(\gamma ^a p_a -mc)\Psi (x^0,{\bf x},c) = (i\hbar \gamma ^0 \partial _0 +
i\hbar \gamma .\nabla -mc)\Psi (x^0,{\bf x},c) =0
\end{equation}
\noindent Here $x^a=(ct,x,y,z)$;  $g_{ab}=diag(+,-,-,-)$;  $\gamma ^a=
(\gamma^0,\gamma)$;  $\gamma=(\gamma^1,\gamma^2,\gamma^3)$ are the
Dirac matrices;  $p_a=i\hbar \partial  /\partial  x^a=i\hbar  \partial
_a$;  $a=0,1,2,3$;  the  summation  is  carried  out  over  the  twice
repeating index; $\Psi =column(\phi,\chi)$; $\phi =column(\phi _1,\phi
_2)$;  $\chi =column(\chi _1,\chi _2)$. The gamma matrices satisfy the
relations \cite{Sch57}, \cite{Ber68}
\begin{equation}
\label{f3}
\begin{array}{c}
\displaystyle
\vspace{1mm}
\gamma ^a \gamma ^b + \gamma ^b \gamma ^a = 2g^{ab}; \ {}
\gamma ^a \gamma ^5 + \gamma ^5 \gamma ^a =0; \\
\displaystyle
(\gamma ^0)^+=\gamma ^0; \ {} (\gamma ^{1,2,3})^+=-(\gamma ^{1,2,3}); \ {}
(\gamma ^0)^2=1; \ {} (\gamma ^{1,2,3})^2=-1; \\
\displaystyle
(\gamma^{0,1,3})^*=\gamma^{0,1,3}; \ {} (\gamma^2)^*=-\gamma^2; \ {}
(\gamma^{0,2})^ {\tt  T}=\gamma^{0,2};  \ {}
(\gamma^{1,3})^ {\tt T}=-\gamma^{1,3}; \\
\displaystyle
\gamma ^5=-i\gamma^0 \gamma ^1 \gamma ^2 \gamma ^3; \ {}  (\gamma ^5)^+=
\gamma^5; \ {} (\gamma^5)^*=\gamma^5; \ {}   (\gamma ^5)^2=1
\end{array}
\end{equation}
\begin{equation}
\label{f4}
\gamma ^0=\left\lgroup\matrix{I&0\cr 0&-I\cr}\right\rgroup;
\gamma ^{1,2,3}=\left\lgroup\matrix{0&\sigma _{x,y,z}\cr -\sigma
_{x,y,z}&0\cr}\right\rgroup; 
\gamma ^5=\left\lgroup\matrix{0&-I\cr -I&0\cr}\right\rgroup
\end{equation}
\begin{equation}
\label{f5}
\sigma _x=\left\lgroup\matrix{0&1\cr1&0\cr}\right\rgroup;
\sigma _y=\left\lgroup\matrix{0&-i\cr i&0\cr}\right\rgroup;
\sigma _z=\left\lgroup\matrix{1&0\cr0&-1\cr}\right\rgroup
\end{equation}
Here I is the unit two dimensional matrix.
\noindent In accordance with Ref.  [1,2],  we  define  the  action  of
operators $P$,  $T$, $C$, $Q$ on the solution $\Psi $ in the following
forms:
\begin{equation}
\label{f6}
P\Psi (x^0,{\bf x},c)=U_P\Psi (x^0,-{\bf x},c);
\end{equation}
\begin{equation}
\label{f7}
T\Psi (x^0,{\bf x},c)=U_T \overline \Psi^ {\tt T} (-x^0,{\bf
x},c)= U_T\gamma ^0 \Psi ^* (-x^0,{\bf x},c);
\end{equation}
\begin{equation}
\label{f8}
C\Psi (x^0,{\bf x},c)=U_C \overline \Psi^ {\tt T}  (x^0,{\bf
x},c)=U_C \gamma ^0 \Psi ^* (x^0,{\bf x},c);
\end{equation}
\begin{equation}
\label{f9}
Q\Psi (x^0,{\bf x},c)=U_Q \Psi (x^0,{\bf x},-c)
\end{equation}
\noindent Here  $U_P$,  $U_T$,  $U_Q$,  $U_C$  are  the  corresponding
matrices of the transformations;  $\overline \Psi =\Psi ^+\gamma^0$;
{\tt T} is the transposition; $*$ is the complex conjugation.
\par We take Eq. (\ref{f2}) and perform the $Q$-inversion in it:
\begin{equation}
\label{f10}
\begin{array}{c}
\displaystyle
Q: \ (\gamma ^a p_a - mc)\Psi (x^0,{\bf x},c)=0 \to \\
\displaystyle
(U_Q\gamma ^a U_Q ^{-1} p_a + mc)U_Q \Psi (x^0,{\bf x},-c)=0
\end{array}
\end{equation}
\noindent 
\begin{sloppypar}
Let Eq.  (\ref{f10}) coincide with the initial Eq.  (\ref{f2}) for the
transformed  function  $\Psi _Q=U_Q \Psi (x^0,{\bf x},-c)$.  It can be
obtained if the matrix $U_Q$ has the following property:
\end{sloppypar}
\begin{equation}
\label{f11}
U_Q\gamma ^a U_Q^{-1}=-\gamma ^a
\end{equation}
\noindent As far as $U_Q\gamma ^a + \gamma ^a U_Q=0$,  it  is  follows
from it and the properties of the $\gamma^ a$-matrices that
\begin{equation}
\label{f12}
U_Q=\lambda \gamma ^5,
\end{equation}
\noindent where $\lambda=(\pm 1,\pm i)$. As a result for attaining the
invariance  of  Dirac  equation  relative  to  the $Q$-inversion it is
necessary to transform the function $\Psi $ following the rule
\begin{equation}
\label{f13}
Q\Psi (x^0,{\bf x},c)=U_Q\Psi (x^0,{\bf x},-c)=\lambda \gamma ^5
\Psi (x^0,{\bf x},-c)
\end{equation}
\noindent We take into account that \cite{Ber68}:
\begin{equation}
\label{f14}
U_P=i\gamma ^0, \ {} U_T=-i\gamma ^0\gamma ^1\gamma ^3, \ {}
U_C=-\gamma ^0\gamma ^2,
\end{equation}
\noindent  use the permutational properties (\ref{f3}) of the $\gamma^a$
- matrices, hold the parameter $\lambda=i$ and make a table of the 
transformations  of  the  function  $\Psi  (x^0,{\bf x},c)$  relative
to the $P$, $T$, $Q$-inversions and the charge conjugation $C$.
\begin{equation}
\label{f15}
\begin{array}{ll}
Berestetski, Lifshits, Pitaevski                                  &
The \ {present} \ {work}                                          \\
\vspace{3mm}                 
C, P, T \ {symmetries}, \ {c \to +c} \ {\cite{Ber68}}:            &  
P, T, Q \ {symmetries}, \ {c \to \pm c}:                          \\ 
P \Psi = i\gamma ^0\Psi (x^0,-{\bf x},c);                         &
P \Psi = i\gamma ^0\Psi (x^0,-{\bf x},c)                          \\
T \Psi = -i\gamma ^1\gamma ^3\Psi ^* (-x^0,{\bf x},c);            &
T \Psi = -i\gamma ^1\gamma ^3\Psi ^* (-x^0,{\bf x},c)             \\
\vspace{3mm}
P T \Psi = \gamma ^0\gamma ^1\gamma ^3\Psi ^* (-x^0,-{\bf x},c);  & 
P T \Psi = \gamma ^0\gamma ^1\gamma ^3\Psi ^* (-x^0,-{\bf x},c);  \\
C P T \Psi = i\gamma ^5 \Psi (-x^0,-{\bf x},c);                   & 
Q \Psi = i\gamma ^5 \Psi (x^0,{\bf x},-c);                        \\
C T \Psi = i\gamma ^1 \gamma ^2 \gamma ^3 \Psi (-x^0,{\bf x},c);  &
P Q \Psi = i\gamma ^1 \gamma ^2 \gamma ^3 \Psi (x^0,-{\bf x},-c); \\
C P \Psi = i\gamma ^0 \gamma ^2 \Psi ^* (x^0,-{\bf x},c);         & 
T Q \Psi = i\gamma ^0 \gamma ^2 \Psi ^* (-x^0, {\bf x},-c);       \\
C \Psi = \gamma ^2 \Psi ^* (x^0,{\bf x},c);                       &
P T Q \Psi = - \gamma ^2 \Psi ^* (-x^0,-{\bf x},-c)
\end{array}
\end{equation}
\begin{sloppypar}
\noindent
One can  see  that  the charge conjugation $C$ with accuracy up to the
phase factor $exp(i\pi)$ corresponds to  the  $PTQ$-transformation  so
that
\begin{equation}
\label{f16}
[C,PTQ]\Psi (x^0,{\bf x},c)=0
\end{equation} 
\end{sloppypar}
\noindent
Similarly, the operations $CPT$, $CT$ and $CP$ correspond to  $Q$, and 
the $PQ$,  $TQ$ compositions.
\par Let  us  consider  in more   details   the   mechanism   of   the
$C\leftrightarrow  PTQ$  correspondence.  Following to \cite{Ber68} we
rewrite the function  $\Psi$  in  explicit  form  for  our  case  when
$c\neq1$, $\hbar\neq1$:
\begin{equation}
\label{f17}
\Psi_{p\sigma}={1\over \sqrt {2p^0}}u_{p\sigma}e^{-{i\over \hbar}p.x}; \
{}
\Psi_{-p-\sigma}={1\over \sqrt         {2p^0}}u_{-p-\sigma}e^{{i\over
\hbar}p.x};
\end{equation}
\begin{equation}
\label{f18}
u_{p\sigma}=\left\lgroup\matrix{\sqrt {p^0+mc} \ {}w\cr \sqrt {p^0-mc}{}
({\bf n}\sigma)w\cr}\right\rgroup;
u_{-p-\sigma}=\left\lgroup\matrix{\sqrt {p^0-mc}({\bf n}\sigma)w'\cr
\sqrt {p^0+mc} \ {}w'\cr}\right\rgroup
\end{equation}
Here $p^0=E/c>0$,  $E=c\sqrt{{\bf  p}^2+m^2c^2}>0$,   $p.x=p^0x^0-{\bf
p}{\bf x}$,  ${\bf n}={\bf p}/p$,  $w^+ w=1$, $w= ({\bf n}\sigma )w'$,
$\overline {u}_p u_p=2mc$,  $\overline {u}_{-p} u_{-p}= -2mc$,  $c>0$.
In  application to functions $\Psi_ {-p-\sigma}$ and $\Psi_ {p\sigma}$
of operators $C$ and $PTQ$ we have:
\begin{equation}
\label{f19}
\begin{array}{c}
\displaystyle
\vspace{2mm}
C\Psi_ {-p-\sigma -E}(x^0,{\bf x},c)=
\gamma^2 \Psi_{-p-\sigma -E}^* (x^0,{\bf x},c)=
\Psi_ {p\sigma E} (x^0,{\bf x},c); \\
\displaystyle
PTQ\Psi_ {p\sigma E}(x^0,{\bf x},c)=
-\gamma^2 \Psi_{p\sigma E}^* 
(-x^0,-{\bf x},-c)= -\Psi_{p\sigma -E} (x^0,{\bf x},-c)
\end{array}  
\end{equation}
We take  into  account  in  the  first expression that $\sigma_y ({\bf
n}\sigma^*)=-({\bf n}\sigma)\sigma_y$,   $-\sigma_y w^*=u'$, $u'^+ u'=
{(\sigma_y w^*)}^+(\sigma_y w^*)= w^T{\sigma_y}^+\sigma_y w^*={(w^+
w)}^*=1$ and in  addition  to  this  in  the  second  expression  that
$p^0=(-E)/(-c)>0$,  ${\bf p}=(-E)(-{\bf v})/c^2>0$. In addition to Eq.
(\ref{f16}) we have in result
\begin{equation}
\label{f20}
C\Psi_{-p-\sigma -E}(x^0,{\bf x},c)=-PTQ\Psi_{p\sigma E}(x^0,{\bf x},c)
\end{equation}
because of function $\Psi_{p\sigma E}(x^0,{\bf x},c)$ 
coincides with the function $\Psi_{p\sigma -E}(x^0,{\bf x},-c)$ due to 
relations $E/c=(-E)/(-c)$, $mc=(-m)(-c)$.

\section{$PTQ$ -symmetry and Dirac equation for a charged particle}
\label{s3}
\par Now  we  consider  the  Dirac equation for a charge particle with
spin 1/2 in electromagnetic field:
\begin{equation}
\label{f21}
(\gamma ^a p_a - mc)\Psi (x,c)=(e/c)\gamma ^a A_a \Psi (x,c)
\end{equation}
\noindent where  $x=(x^0,{\bf  x})$,  $e$ is the charge of a particle,
$A^a=(A^0,{\bf A})$ is  the  4-potential,  $\gamma^a  A_a=\gamma^0
A^0-\gamma .{\bf A}$,  $\gamma =(\gamma^1, \gamma^2,\gamma^3)$. Let
us subject the equation (\ref{f21}) to the $PTQ$ transformation taking
into  account the formulas (\ref{f6}),  (\ref{f7}),  (\ref{f9}) on the
assumption that an  electrical  charge  is  a  scalar;  the  vector  -
potential is a polar vector relative to the replacements both ${\bf x}
\to -{\bf x}$,  and $t \to -t$,  and $c \to -c$ \cite{Kot92}.  Let  us
carry out the transformation $Q$, then $T$, then $P$:

\begin{equation}
\label{f22}
\begin{array}{c}
\displaystyle
Q(x^0,{\bf x},c,e,A^0,{\bf A})=(x^0,{\bf x},-c,e,A^0,{\bf A}); \\
\displaystyle
\vspace{1mm}
Q\Psi (x,c)=\Psi _Q (x,-c)=U_Q\Psi (x^0,{\bf x},-c); \\
\displaystyle
Q: \ (\gamma ^a p_a - mc)\Psi  = (e/c)\gamma ^a A_a \Psi \to \\
\displaystyle
(i\hbar U_Q\gamma ^0 U_Q ^{-1} \partial _0
+ i\hbar U_Q \gamma U_Q ^{-1}.\nabla 
+ mc)U_Q \Psi (x^0,{\bf x},-c)= \\
\displaystyle
(-e/c)(U_Q\gamma ^0 U_Q ^{-1} A^0 - U_Q\gamma U_Q ^{-1} {\bf A})
U_Q \Psi (x^0,{\bf x},-c) \to \\
\displaystyle
(\gamma ^a p_a - mc)\Psi _Q= (-e/c)\gamma ^a A_a \Psi _Q
\end{array}
\end{equation}
\begin{equation}
\label{f23}
\begin{array}{c}
\displaystyle
T(x^0,{\bf x},c,e,A^0,{\bf A})=(-x^0,{\bf x},c,e,A^0,-{\bf A}); \\
\displaystyle
\vspace{1mm}
T\Psi _Q (x,-c)=\Psi_{TQ} (-x^0,{\bf x},-c)=
U_T\overline {\Psi }_Q ^{\tt T}
(-x^0,{\bf x},-c); \\
\displaystyle
T: \ (\gamma ^a p_a - mc)\Psi _Q=(-e/c)\gamma ^a A_a\Psi _Q \to \\
\displaystyle
(i\hbar U_T\gamma ^{0 \tt T} U_T ^{-1} \partial _0
- i\hbar U_T \gamma ^{\tt T} U_T ^{-1}.\nabla
- mc)U_T \overline {\Psi }_Q ^{\tt T} (-x^0,{\bf x},-c)= \\
\displaystyle
(-e/c)(U_T\gamma ^{0   \tt T}  U_T  ^{-1}  A^0  +  U_T\gamma
^{\tt T}U_T ^{-1} {\bf A})
U_T\overline {\Psi }_Q ^{\tt T} (-x^0,{\bf x},-c) \to \\
\displaystyle
(\gamma ^a p_a - mc)\Psi _{TQ}= (-e/c)\gamma ^a A_a \Psi _{TQ}
\end{array}
\end{equation}
\begin{equation}
\label{f24}
\begin{array}{c}
\displaystyle
P(x^0,{\bf x},c,e,A^0,{\bf A})=(x^0,-{\bf x},c,e,A^0,-{\bf A}); \\
\displaystyle
\vspace{1mm}
P\Psi _{TQ}(-x^0,{\bf  x},-c)=\Psi  _{PTQ}(-x,-c)= 
U_P\Psi_{TQ}(-x^0,-{\bf x},-c); \\
\displaystyle
P: \ (\gamma ^a p_a - mc)\Psi _{TQ}=(-e/c)\gamma ^a A_a
\Psi _{TQ} \to \\
\displaystyle
(i\hbar U_P\gamma ^0 U_P ^{-1} \partial _0
- i\hbar U_P \gamma U_P ^{-1}.\nabla
- mc)U_P \Psi _{TQ}(-x^0,-{\bf x},-c)= \\
\displaystyle
(-e/c)(U_P\gamma ^0 U_P ^{-1} A^0 + U_P\gamma U_P ^{-1} {\bf A})
U_P \Psi _{TQ}(-x^0,-{\bf x},-c) \to \\
\displaystyle
(\gamma ^a p_a - mc)\Psi _{PTQ}= (-e/c)\gamma ^a A_a \Psi _{PTQ}
\end{array}
\end{equation}
\noindent Here matrices $U$ satisfy the conditions
\begin{equation}
\label{f25}
\begin{array}{c}
U_Q\gamma ^0 U_Q^{-1}=-\gamma ^0, \ {}
U_Q\gamma U_Q^{-1}=-\gamma ; \\
U_T(\gamma ^0)^{\tt T}U_Q^{-1}=\gamma ^0, \ {}
U_T\gamma ^{\tt T}U_T^{-1}=-\gamma;  \\
U_P\gamma ^0 U_P^{-1}=\gamma ^0, \ {} U_P\gamma U_P^{-1}=-\gamma ,
\end{array}
\end{equation}
which define their explicit forms (\ref{f12}) and (\ref{f14}).  On the
basis  of  the  formulae  (\ref{f15})  and  taking  into account $\Psi
_{PTQ}=-\gamma^2\Psi ^* (-x^0,-{\bf x},-c)$, we can write
\begin{equation}
\label{f26}
(\gamma ^a p_a - mc)\gamma ^2\Psi ^* (-x,-c)=
(-e/c)\gamma ^a A_a \gamma ^2\Psi ^* (-x,-c)
\end{equation}
\noindent where  $-x=(-x^0,-{\bf  x})$.  Similarly   to   the   charge
conjugation,   the   equation  received  coincides  with  initial  Eq.
(\ref{f21}) for the electric charge $-e$ and the transformed  function
$-\gamma ^2\Psi ^*(-x,-c)$.  In accordance with formula (\ref{f19}) it
is possible to admit that Eq.  (\ref{f26}) describes a  particle  with
the  charge  $-e$,  momentum $p=(p^0>0,{\bf p}>0)$ and negative energy
$E<0$ $(p^0=(-E)/(-c)>0,{\bf p}=(-E)(-{\bf v})/c^2>0)$. The energy gap
of $2E$ is present between $c>0$ and $c<0$-states.
\begin{sloppypar}  Thus,  the  charge  conjugation   $C$   puts   into
correspondence  the antiparticle with characteristics $(-e,m,-p,-E,c)$
to   the   particle   with    characteristics    $(e,m,p,E,c)$.    The
$PTQ$-composition   puts   into   correspondence   the  particle  with
characteristics $(-e,m,p,-E,-c)$ to the particle with  characteristics
$(e,m,p,E,c)$.   From   here   it  is  seen  that  the  particle  with
characteristics $(-e,m,p,-E,-c)$ may  be  the  redefined  antiparticle
with respect to the initial particle from Eq. (\ref{f21}).
\end{sloppypar}
\par As  in  the  case  of  the  $C$-conjugation   \cite{Ber68},   the
$PTQ$-composition is the symmetry transformation of Dirac equation for
a charged particle (\ref{f21}) if the 4-potential  of  electromagnetic
field $A$ is transformed following the rule $QPT(A)=(-A^0,-{\bf A})$.

\section{Conclusion}
\label{s4}
\par The inversion of the speed of light $Q:  x^0 \to x^0, {\bf x} \to
{\bf  x},c  \to -c$ was considered in Dirac equation.  As a result the
charge conjugation $C$ in quantum theory may  be  interpreted  as  the
consequence  of  the  Dirac  equation  symmetry  with  respect  to the
discrete transformations $P$,  $T$,  $Q$  reflecting  the  fundamental
properties  of  space,  time  and  speed  of light:  $C\Psi_{-p-\sigma
-E}(x^0,{\bf x},c)=-PTQ\Psi_{p\sigma E} (x^0,{\bf x},c)$.  We may note
the following consequences of the $PTQ$ -symmetry.
\begin{itemize}
\item
It is not improbable that the  $V^5(x^0,{\bf  x},c)$-space  exists  in
which   there   are  two  hyperplanes  with  $c=+3.10^{10}$  ám/s  and
$c=-3.10^{10}$ ám/s \cite{Kot92}.  In this space the  $c<0$-hyperplane
with  the reversal "time" $x^0 \to -x^0$ and with inverted space ${\bf
x} \to -{\bf x}$ may be interpreted as some antiworld  separated  from
the $c>0$ -world by the energy gap of $2E$.  That may be the redefined
electron-photon Dirac vacuum.  In such vacuum a particle with negative
energy is characterized by its placing on another hyperplane of common
5-dimensional physical space of events.
\item It is impossible to distinguish between the realization  of  the
4-dimensional physical event space on the "$+c$-hyperplane" and on the
"$-c$-hyperplane". From the point of view of an observer placed on the
"$+c$-hyperplane",   the   "$-c$-vacuum"   is   invisible;   and   the
"$+c$-vacuum" is invisible for an observer on the "$-c$-hyperplane".
\item The  additional  quantum  number,  which  is  connected with the
$Q$-parity being kept, may exist in quantum interactions.
\item The  alternative  interpretation of the $K^+$ meson decay scheme
${K^+}_{\pi 3} \to \pi^+ + \pi^+ + \pi^-$;  ${K^+}_{\pi 2} \to \pi^+ +
\pi^0$. The scheme may be explained if the $K^+$ and $\pi$ mesons have
the negative $Q$-parity.
\end{itemize}

\end{document}